\documentclass[a4paper,12pt]{article}
\usepackage{graphicx}
\usepackage{amssymb}
\usepackage{amsmath}
\usepackage[doublespacing]{setspace}
\usepackage[a4paper,portrait,twocolumn=false,includeheadfoot,mag=1000,dvips]{geometry}
\usepackage[title]{appendix}
\usepackage{caption}
\usepackage{float}
\usepackage{color}
\usepackage{ulem}
\usepackage{mathrsfs}
\usepackage{type1cm}
\usepackage{marvosym}
\usepackage{amssymb}
\usepackage{wasysym}
\usepackage{subfigure}
\usepackage{multirow}
\usepackage{booktabs}
\usepackage[utf8]{inputenc}

\makeatletter

\newcommand{\Rmnum}[1]{\expandafter\@slowromancap\romannumeral #1@}
\makeatother

\numberwithin{equation}{section}

\newcommand{\sanhao}{\fontsize{12pt}{16pt}\selectfont}

\newcommand{\xiaosihao}{\fontsize{13pt}{\baselineskip}\selectfont}

\begin{document} \xiaosihao
\newtheorem{The}{Theorem}
\newtheorem{exam}{Example}
\newtheorem{lem}{Lemma}
\newtheorem{de}{Definition}
\newtheorem{prop}{Proposition}
\newtheorem{cor}{Corollary}
\newtheorem{hyp}{Hypothesis}
\newtheorem{rem}{Remark}
\pagestyle{plain}

\setlength{\baselineskip}{20pt}
\setlength{\parskip}{0.4\baselineskip}

\clearpage
\begin{center}

\xiaosihao \textbf{THE DINEGENTROPY OF DIAGNOSTIC TESTS}



{\normalsize Nozer D. Singpurwalla}\\[-10pt]
{\normalsize The George Washington University, Washington, D.C., USA}\\[-10pt]
{\normalsize and}\\[-13pt]
{\normalsize Boya Lai}\\[-13pt]
{\normalsize The City University of Hong Kong, Hong Kong}\\
{\normalsize (August 2020)}

\end{center}

\renewcommand{\abstractname}{{\large Abstract}}
\begin{abstract}
{\sanhao
	Diagnostic testing is germane to a variety of scenarios in medicine, pandemic tracking, threat detection, and signal processing. This is an expository paper with some original results. Here we first set up a mathematical architecture for diagnostics, and explore its probabilistic underpinnings. Doing so enables us to develop new metrics for assessing the efficacy of different kinds of diagnostic tests, and for solving a long standing open problem  in diagnostics, namely, comparing tests when their receiver operating characteristic curves cross. The first is done by introducing the notion of what we call, a \textit{\textbf{Gini Coefiicient}}; the second by invoking the information theoretic notion of \textit{\textbf{dinegentropy}}. Taken together, these may be seen a contribution to the state of the art of diagnostics. \\[-5pt]
    
    The spirit of our work could also be relevant to the much discussed topic of \textit{\textbf{batch testing}}, where each batch is defined by the partitioning strategy used to create it. However this possibility has not been explored here in any detail. Rather, we invite the attention of other researchers to investigate this idea, as future work.\\ [10pt]
}

\end{abstract}
\addtocounter{section}{-1}
\newpage
\section{Background.}
Diagnostic testing has played a key role in medicine, threat detection, machine learning, supervised classification, signal  processing, verification of treaties, alerts for tornadoes, tsunamis, and earthquakes, pandemic tracking, and a host of other activities. However, diagnostic tests are not perfect -- they are prone to misdiagnosis and false alarms. This is due to the random nature of the attributes they monitor. The goal of this paper is to develop approaches for assessing the efficacy of a test, and for comparing the performance of two different tests, when their \textit{receiver operating characteristics} cross. The latter has been a long standing open problem in diagnostics.

With the above in mind, we introduce some fundamentals, followed by an architecture for discussing the underlying mathematics. The material of Section 1 is mainly expository; however, what is offered are new perspectives. They enable one to explore the anatomy of diagnostic tests and to provide a deeper appreciation of their probabilistic structure. Section 2 describes relationships between two key parameters in diagnostics, \textit{sensitivity} and \textit{specificity}. Whereas these relationships are known, the manner in which we interpret them paves the path for much that follows. Section 3 pertains to a key metric in diagnostics, \textit{the area under the curve}. It is a stepping stone to that of Section 4, which is the thrust of this paper. The materials of Sections 2 and 3 encapsulate the flow of ideas leading to Section 4, and serve the purpose of bringing together some scattered results in this arena.

Section 4 brings to closure, the material in the preceding sections by introducing two new notions, the \textit{Gini Coefficient}, (which is not the same as the Gini Index), and \textit{dinegentropy}. The first is able to assess the efficacy of any diagnostic test by comparing it against a completely random test; i.e. a test whose diagnosis is based on the outcome of the flip a coin. The information-theoretic notion of dinegentropy enables the comparison of tests whose receiver operating characteristic curves cross. Doing so enables us to resolve a long standing problem in diagnostic testing.
\section{Preliminaries.}

Assume a single disease and suppose the disease spawns a measurement $Z$, that is compared to a threshold $T$. An individual is classified $D$ (diseased) if $Z>T$; otherwise the individual is classified $N$ (normal). The simplicity of this mechanism vanishes when one or both $Z$ and $T$ are random, as perceived by a diagnostician $\mathscr{D}$. With T fixed at t, and $Z$ random, $\mathscr{D}$ will experience errors of misclassification and false alarms. Similarly with T random and $Z$ fixed. 

With $T$ preselected at $t$, the probability of correctly classifying a diseased individual is known as the test \textit{sensitivity}, $S_e(t)$. Test \textit{specificity} is the probability of correctly classifying a normal individual as such; it is denoted $S_p(t)$. Ideally, one wants both $S_e(t)$ and $S_p(t)$ to be close to one, but for this to happen, the distribution of $Z$ for the case $D$ should not overlap with its distribution for the case $N$; also, $t$ should be selected in a region where there is no overlap. Otherwise an increase in $S_e(t)$ will cause a decrease in $S_p(t)$ and vice-versa. Test sensitivity and specificity are \textit{adversarial} parameters.

Let $F_0(F_1)$ be the distribution of $Z$ for individuals in the class $N(D)$, and let $Z_0(Z_1)$ denote their corresponding random variables. Suppose that the supports of $Z_0$ and $Z_1$ overlap, but that $Z_1\overset{st}{>}Z_0$; i.e. for all $z$, $\bar{F}_1(z)\geq \bar{F}_0(z)$ where $\bar{F}=1-F$. In the archetypal scenario, $T$ is precisely set to $t$, and $F_0$ and $F_1$ are, assumed absolutely continuous and fully specified, with $f_0$ and $f_1$ denoting their probability densities.

Observe that $S_e(t)=1-F_1(t)=\bar{F}_1(t)$, and $S_p(t)=F_0(t)$, so that $\bar{F}_0(t)=1-S_p(t)$.  $\bar{F}_0(t)$ is known as the \textit{false positive probability} (FPP). Increasing $S_e(t)$ by lowering $t$, decreases $S_p(t)$; that is, increasing sensitivity also increases FPP. The adversarial nature of sensitivity and specificity is encapsulated via the \textit{receiver operating characteristic curve} (ROC), which is a plot of ($1-S_p(t)$) versus $S_e(t)$, over all values of $t\geq 0$; see Figure 1.1.
\begin{figure}[!ht]
  \centering
  \includegraphics[scale=0.55]{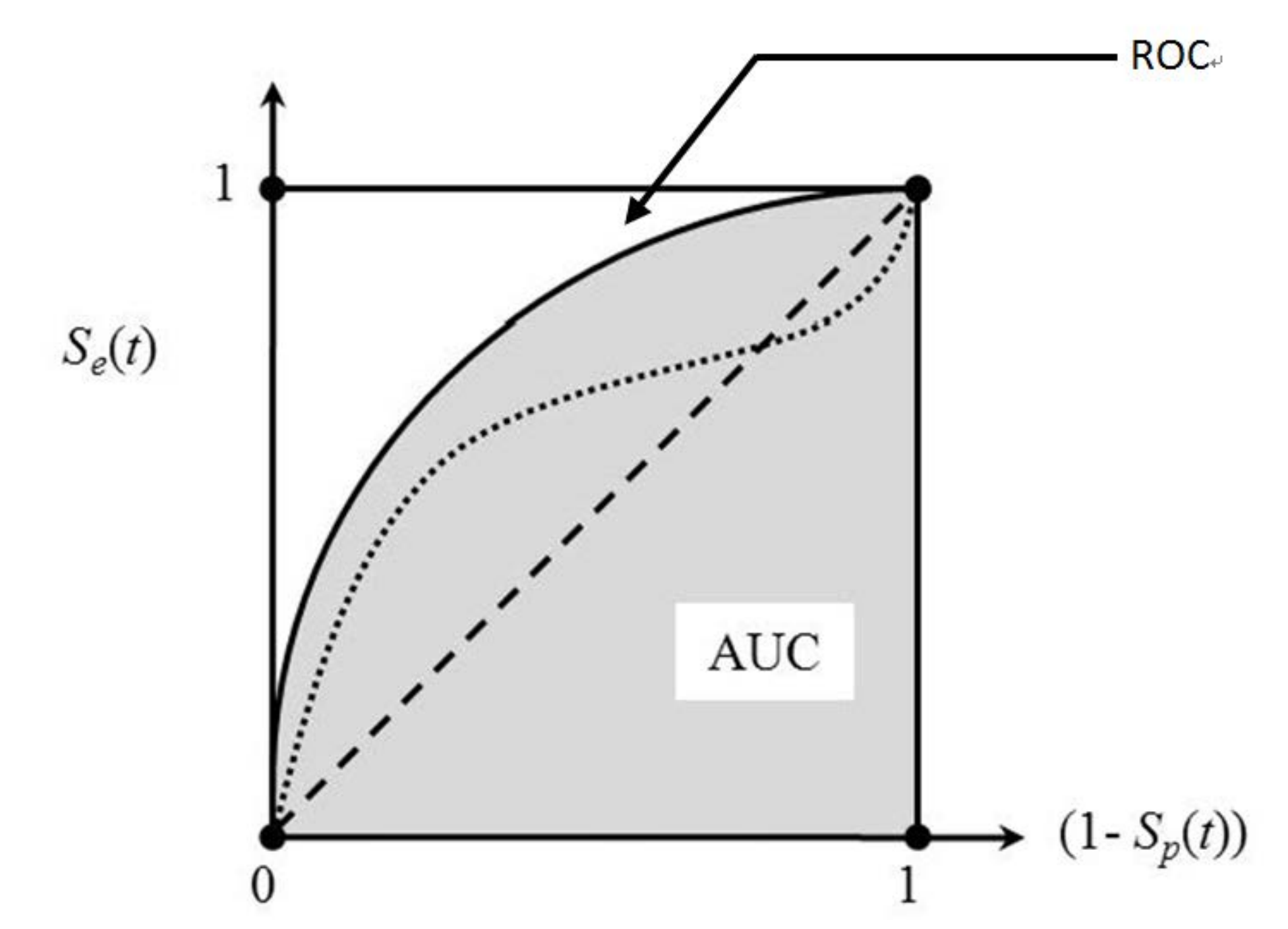}\\
  \caption{The Receiver Operating Characteristic Curve.}
\end{figure}

The ROC is monotonically increasing, continuous, and everywhere differentiable. It is concave in ($1-S_p(t)$), if for all $z$, $F_0(z)\geq F_1(z)$. The ROC will have a convex segment whenever $F_1(z)>F_0(z)$; see the dotted curve of Figure 1.1.

\section{Relationship Between $S_e(t)$ and $S_p(t)$}

For any $t\geq 0$, sensitivity and specificity are related as $S_e(t)=1-F_1F_0^{-1}(S_p(t))$.
To see this, let $X=1-S_p(t)$, and $Y=S_e(t)$. Then $t=S_p^{-1}(1-X)$, so that $y=S_e[S_p^{-1}(1-X)]=1-F_1[S_p^{-1}(1-X)]$, because $S_e(t)=1-F_1(t)$.\\
Since $S_p(t)=F_0(t)$, $S_p^{-1}(t)=F_0^{-1}(t)$, and thus $Y=1-F_1[F_0^{-1}(1-X)]$.  

This result paves the path for developing the information-theoretic distance measure of Section 4. To appreciate this, suppose that $F_0(t)\geq F_1(t)$, for all $t\geq 0$, and examine Figure 2.1 which is a graphic of the spirit of the above result.

\setcounter{figure}{0}
\begin{figure}[!ht]
  \centering
  \includegraphics[scale=0.85]{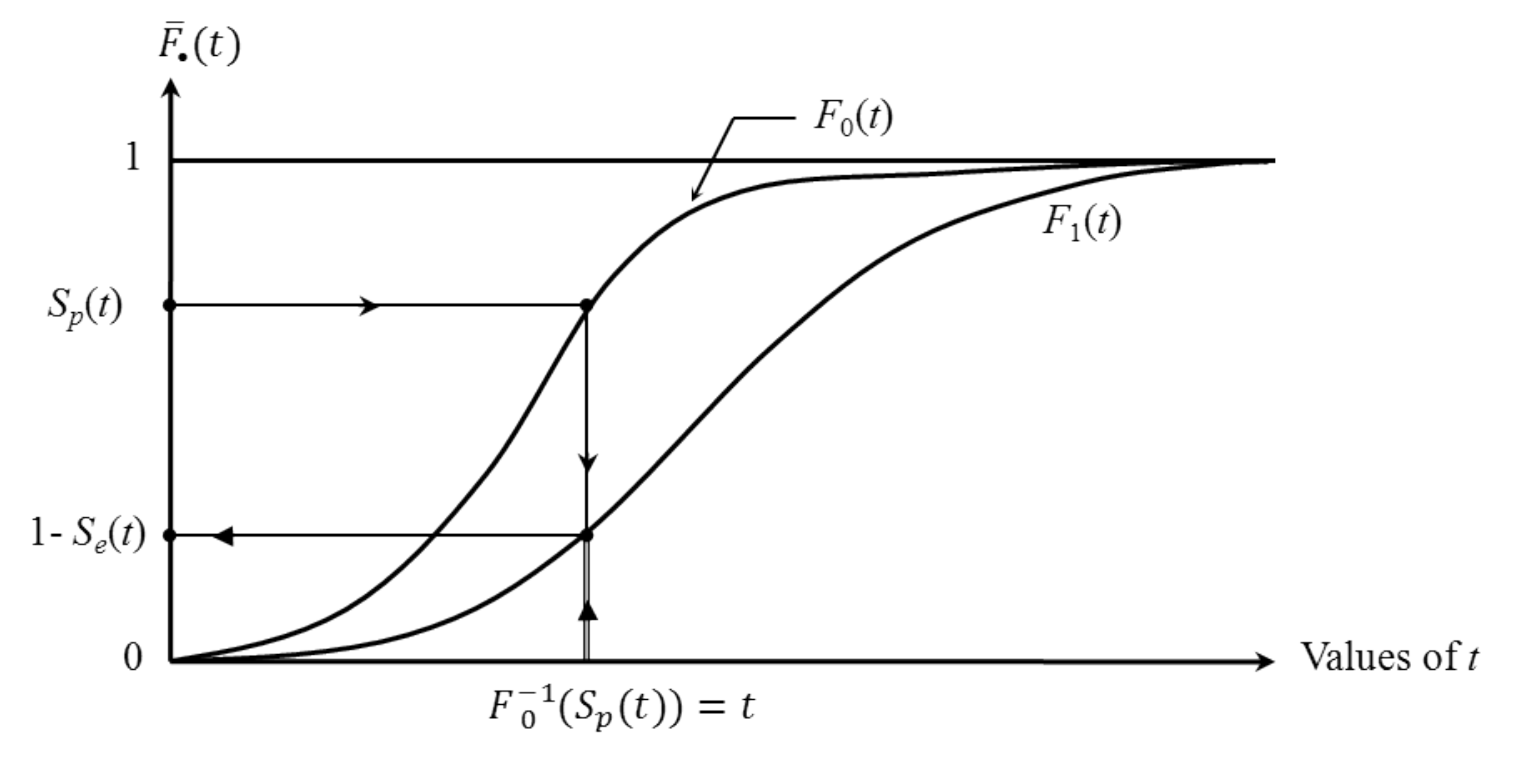}\\
  \caption{Relationship Between Specificity and Sensitivity.}
\end{figure}

Thus corresponding to every t, there exists an $S_e(t)$ and an $S_p(t)>1-S_e(t)$, whenever $F_0(t)>F_1(t)$; when $F_0(t)=F_1(t)$, $S_p(t)=1-S_e(t)$. Figure 2.2 which is a plot of $F_1(t)$ versus $F_0(t)$, illustrates this relationship. Figure 2.3 is a clockwise rotation by $180^{\circ}$ of Figure 2.2, and is the ROC curve.
\begin{figure*}[h]
\centering
\begin{minipage}[h]{0.5\linewidth}
\includegraphics[scale=0.8]{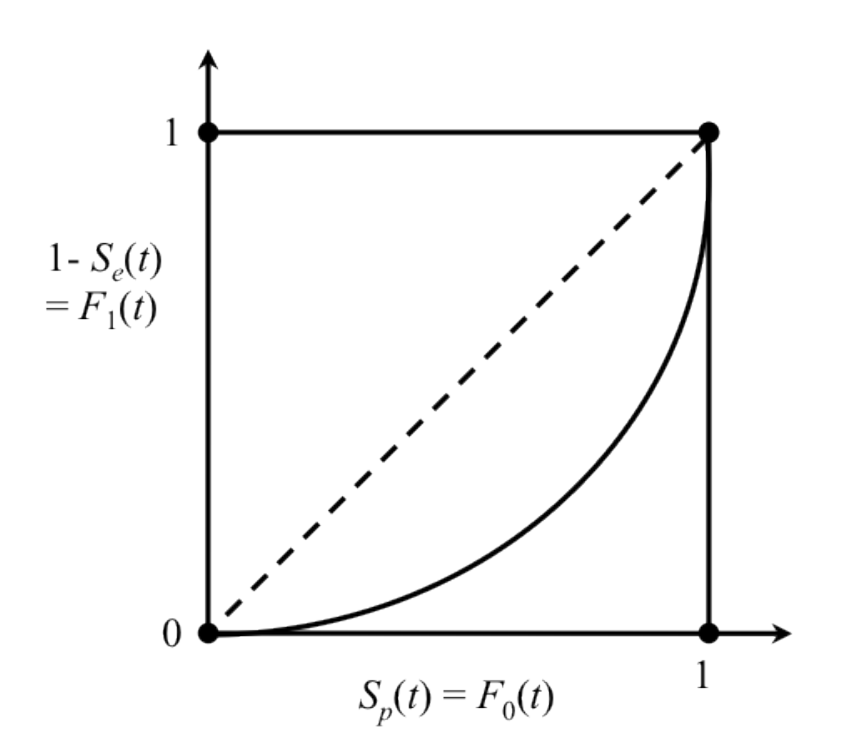}
\caption{Plot of $F_1(t)$ versus $F_0(t)$}
\label{fig:side:a}
\end{minipage}%
\begin{minipage}[h]{0.5\linewidth}
\includegraphics[scale=0.39]{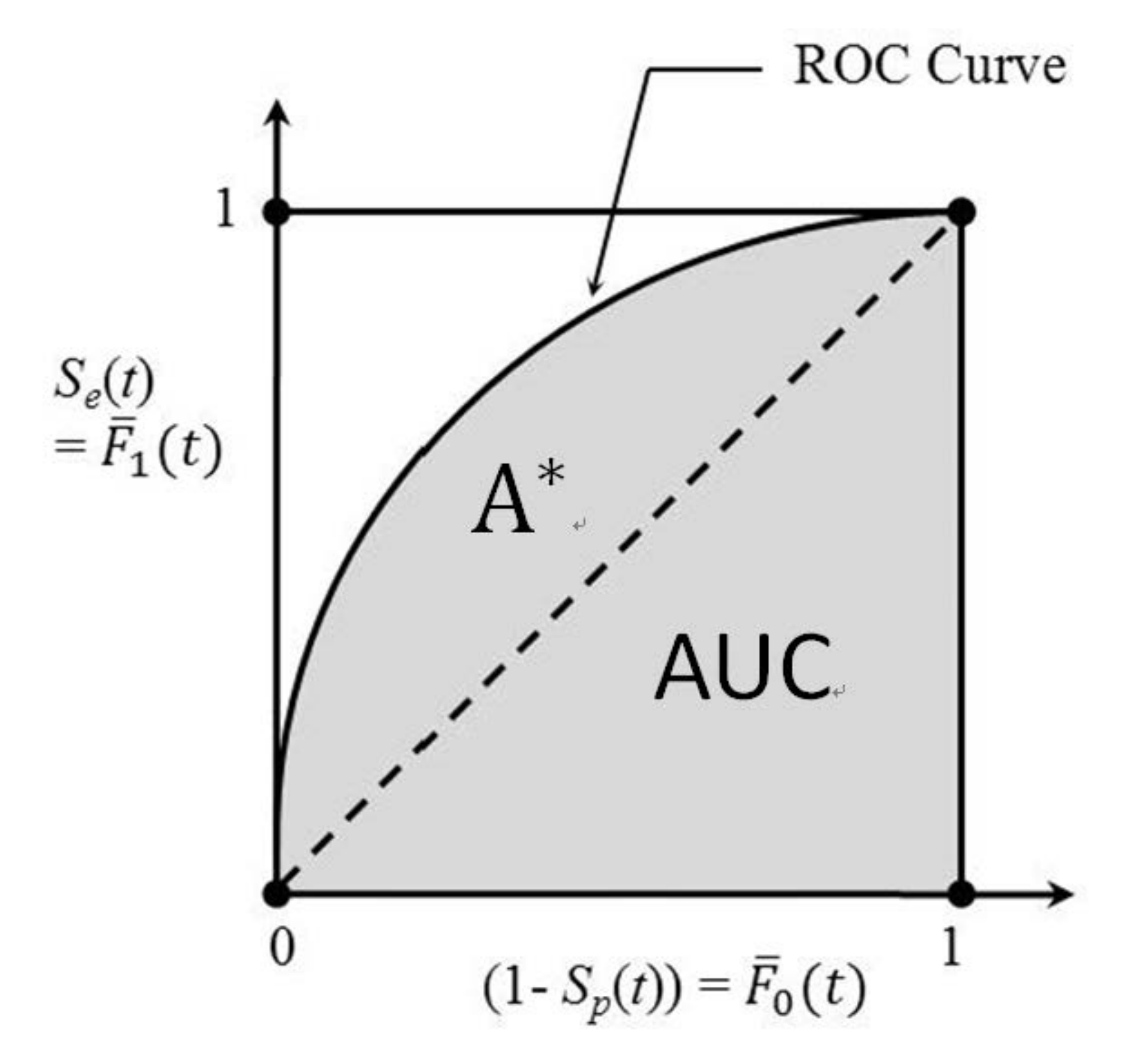}
\caption{The ROC curve}
\label{fig:side:b}
\end{minipage}
\end{figure*}

The essence of Figure 2.3 is that the ROC curve encapsulates a relationship between $F_0(t)$ and $F_1(t)$ over values of $t\geq 0$. It is a profile of the difference between the ordinates of $\bar{F}_0(t)$ and $\bar{F}_1(t)$, and as such is a distance metric. Recognition of this distance metric is the key to the material of Section 4 on dinegentropy, which is based on the Kullback-leibler distance.

\section{Area Under the Curve.}

With $F_0$ and $F_1$ specified, the ROC curve has been used to compare the overall efficacy of two or more diagnostic tests (or sensors), each having their own classification thresholds, over a range of applications. The ROC can also be used to explore the consequences of varying the threshold of a single sensor. Such comparisons can be done in one of several  possible ways, and are useful for selecting tests or test instruments. 

An omnibus metric for comparison is the \textit{area under the ROC curve} -- AUC -- see Figures 1.1 and 2.3. The bigger the AUC, the better the test, provided that the ROC curves do not cross. When the curves cross, using the AUC as a metric can be misleading. The disadvantage of crossing ROC curves has been a challenge, and alternatives to the AUC have been proposed by many; see for example, Gigliarano, Figini, and Muliere (2014). This paper proposes another alternative, provides a rationale for it, and demonstrates its merit.

Figure 2.3 suggests that the AUC can take values between 0 and 1, with 1 corresponding to a perfect diagnosis. This is because the ROC curve encapsulates the relationship between test sensitivity and false positive probability, with the feature that the steeper the ROC curve, the larger the test sensitivity and the smaller the FPP.

When $F_0(t)=F_1(t)$ for all $t\geq0$, the AUC will be $.5$. This is the case of a random diagnosis; that is a test for which the diagnosis is based on the outcome of a flip of an unbiased coin. An AUC of 0 is the consequence of $S_e(t)=0$ for all $t\geq0$ and ($1-S_p(t)$) taking all values between 0 and 1. Tests having an AUC of 0 are worse than a random test, because the latter will be correct at least half of the time. In general a random test is superior to any test whose ROC curve lies below the diagonal line of Figure 2.3.

\section{Test Efficacy: Gini Coefficient and Dinegentropy}

The essence of this paper is Section 4. Here we proposed a solution to the problem of comparing the efficacy of tests whose ROC's cross, and demonstrate its viability. Our solution is based on information theoretic ideas. To motivate our approach we introduce the notion of a \textit{Gini Coefficient}, which is like the Gini Index, but is \underline{not} the Gini Index.

The graphic of Figure 2.3 suggests that the ROC curve is a continuous, non-decreasing function ranging from 0 to 1. Thus the ROC curve can be seen as a distribution function with support on [0,1]. Similarly, the dashed diagonal line of Figure 2.3 can also be seen as the distribution function of a uniform [0,1] distribution. The disparity between these distributions is the region marked $A^{\ast}$. Then, the \textbf{Gini Coefficient} generated by $F_0$ and $F_1$ is defined as $G_{F_0,F_1}=2A^{\ast}$. Since the $AUC=A^{\ast}+.5$, we have
\begin{eqnarray}
G_{F_0,F_1}=2(AUC)-1.
\end{eqnarray}
The Gini Coefficient encompasses three distributions, $F_0$, $F_1$, and the uniform, with $F_0$ and $F_1$ conglomerated as an ROC curve. It encapsulates the idea of the extent to which any diagnostic test is superior to a purely random test. Whereas the Gini Coefficient serves well as an omnibus measure of a diagnostic test's superiority over a random test, it does not obviate the difficulty of comparing tests whose ROC curves cross. This is to be expected because the Gini Coefficient is simply a transformation of the AUC. However, the Gini Coefficient being a measure of the disparity between distribution functions, one being the composition of $F_0$ and $F_1$, and the other being a uniform on [0,1], motivates the development of a measure for comparing two diagnostic tests, irrespective of whether their ROC curves cross or not.

\subsection{\textbf{\textit{Information-Theoretic Considerations in Diagnostics}}}

The earliest work in the arena of using information-theoretic measures in diagnostics is due to Metz, Goodman, and Rossman (1973), whose development focussed on any two points on a single ROC curve, or any two points, one on each of two ROC curves. Their approach is therefore not encompassing, it being restricted to two points. Lee(1999) obtains the \textit{Kullback-Leibler distance} between a two point prior distribution (for non-diseased and diseased individuals), and the corresponding two point posterior distribution. Lee's approach does not compare the entire ROC curve, and is therefore also restricted.

The Jeffreys-Good distance, termed \textit{dinegentropy} by Good (1989), is a measure of concentration for compapring two distribution functions. It is the sum of two Kullback-Leibler distances of two distributions, one, (a reference distribution), and the other an alternate. In the context of diagnostics, the two distributions in question are the ROC curve (or its reflection) and the uniform. Different ROC curves result in different dinegentropies, making it possible to compare two diagnostic tests via their dinegentropies. Theorem 4.1 gives the dinegentropy of an ROC curve. \\[30pt]
\textbf{Theorem 4.1.} \textit{The dinegentropy of an ROC curve with $f_0$ and $f_1$ as the underlying probability densities of the non-diseased and the diseased classes, respectively, is}
\begin{eqnarray*}
\int_0^{\infty}[f_0(t)-f_1(t)]\cdot \text{log}_2 \left( \frac{f_0(t)}{f_1(t)} \right)dt.
\end{eqnarray*}\\[5pt]
\textbf{\textit{Proof}}. It is convenient to work with the reflected ROC curve prescribed as
\begin{eqnarray*}
1-S_e(t)=F_1F_0^{-1}(S_p(t)), \qquad 0\leq S_p(t)\leq 1.
\end{eqnarray*}
For $\bar{F}_1(t)\geq \bar{F}_0(t)$, for all $t\geq 0$, the above is a convex non-decreasing function of $S_p(t)$ whose derivative $(F_1F_0^{-1})'(S_p(t))$ will exist at all $S_p(t)$ if $F_1F_0^{-1}$ is absolutely continuous. Since
\begin{eqnarray*}
(F_1F_0^{-1})'(S_p(t))=F_1'(F_0^{-1}(S_p(t)))\cdot (F_0^{-1})'(S_p(t)),
\end{eqnarray*}
the probability density function generated by $F_1F_0^{-1}(\cdot)$ at $S_p(t)$ is
\begin{eqnarray*}
\frac{f_1(F_0^{-1}S_p(t))}{f_0(F_0^{-1}S_p(t))}=\frac{f_1(t)}{f_0(t)},
\end{eqnarray*}
the \textit{likelihood ratio} at $t$; recall that $S_p(t)=F_0(t)$.

\bigskip

Using $F_1F_0^{-1}$ as the reference distribution, the Kullback-Leibler (K-L) distance is
\begin{eqnarray*}
\int_0^{\infty}\text{log}_2\left(\frac{f_0(t)}{f_1(t)}\right)f_0(t)dt=\int_0^{\infty}\text{log}_2(f_0(t))\cdot f_0(t)dt-\int_0^{\infty}\text{log}_2(f_1(t))\cdot f_0(t)dt.
\end{eqnarray*}
In getting to the above we use the fact that
\begin{eqnarray*}
\frac{d}{dt}S_p(t)=\frac{d}{dt}F_0(t)=f_0(t).
\end{eqnarray*}

Next, we use the diagonal ROC as a reference distribution to obtain the K-L distance as
\begin{eqnarray*}
\int_0^{\infty}\text{log}_2\left(\frac{f_1(t)}{f_0(t)}\right)\cdot \frac{f_1(t)}{f_0(t)}\cdot f_0(t)dt=\int_0^{\infty}log_2(f_1(t))\cdot f_1(t)dt-\int_0^{\infty}log_2(f_0(t))\cdot f_1(t)dt.
\end{eqnarray*}

The statement of the theorem now follows because dinegentropy is the sum of the two K-L distances. \qquad \qquad \qquad \qquad \qquad \qquad \qquad \qquad \qquad \qquad  $\parallel$

\subsubsection{Illustrating the Efficacy of Theorem 4.1}

As proof of principle of Theorem 4.1 for discriminating between two ROC curves with the same AUC, and also to address the matter as to whether an ROC curve with a large dinegentropy is superior to one with a smaller dinegentropy, we look at the following example.

Consider two beta distributions, one having parameters (1,3), and the other having
parameters (2,6). Their distribution functions behave like two ROC curves on the
interval [0,1]. The AUC's corresponding to both of these distributions is 3/4. Thus
using the AUC, as a metric for comparing these two ROC curves will not be meaningful.
Figure 4.1 shows the two ROC curves in question, superimposed on each other, and also
the diagonal ROC curve.

\setcounter{figure}{0}
\begin{figure}[h]
  \centering
  \includegraphics[scale=0.8]{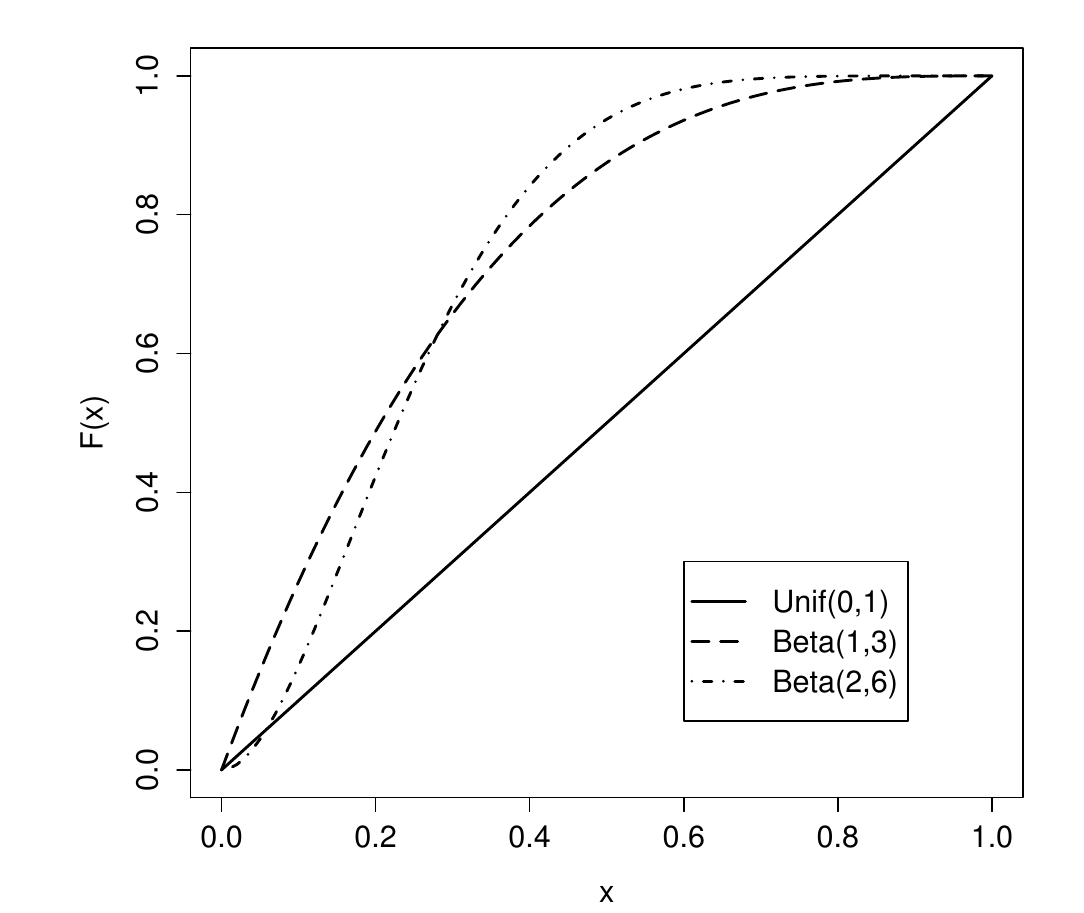}\\
  \caption{Crossing ROC curves with a Common AUC.}
\end{figure}

Since we do not know the two probability density fucntions $f_0$ and $f_1$ of Theorem 4.1 which generated the two ROC curves of Figure 4.1, but we do know the probability density function generated by beta distributions as the ROC, we can directly obtain the K-L distance of each ROC curve using first the uniform distribution as the reference distribution, and then reversing the roles to obtain the dinegentropy.

For the ROC curve corresponding to the beta(1,3) distribution, the probability density is $3(1-x)^2$, for $0\leq x\leq 1$. For the uniform on [0,1] the probability density is 1, for $0\leq x\leq 1$. Thus the K-L distance with the uniform on [0,1] as the reference distribution is
\begin{eqnarray*}
\int_0^1 3(1-x)^2 \text{log}_2(3(1-x)^2)dx=0.623166.
\end{eqnarray*}
Repeating the above exercise but now using the beta(1,3) distribution as the reference, gives the Kullback-Leibler distance as
\begin{eqnarray*}
\int_0^1 log_2(\frac{1}{3(1-x)^2})dx=1.30043.
\end{eqnarray*}
The dinegentropy of the beta(1,3) as the ROC curve is therefore $0.623166+1.30043 =
1.923596$.

\begin{figure}[h]
  \centering
  \includegraphics[scale=0.8]{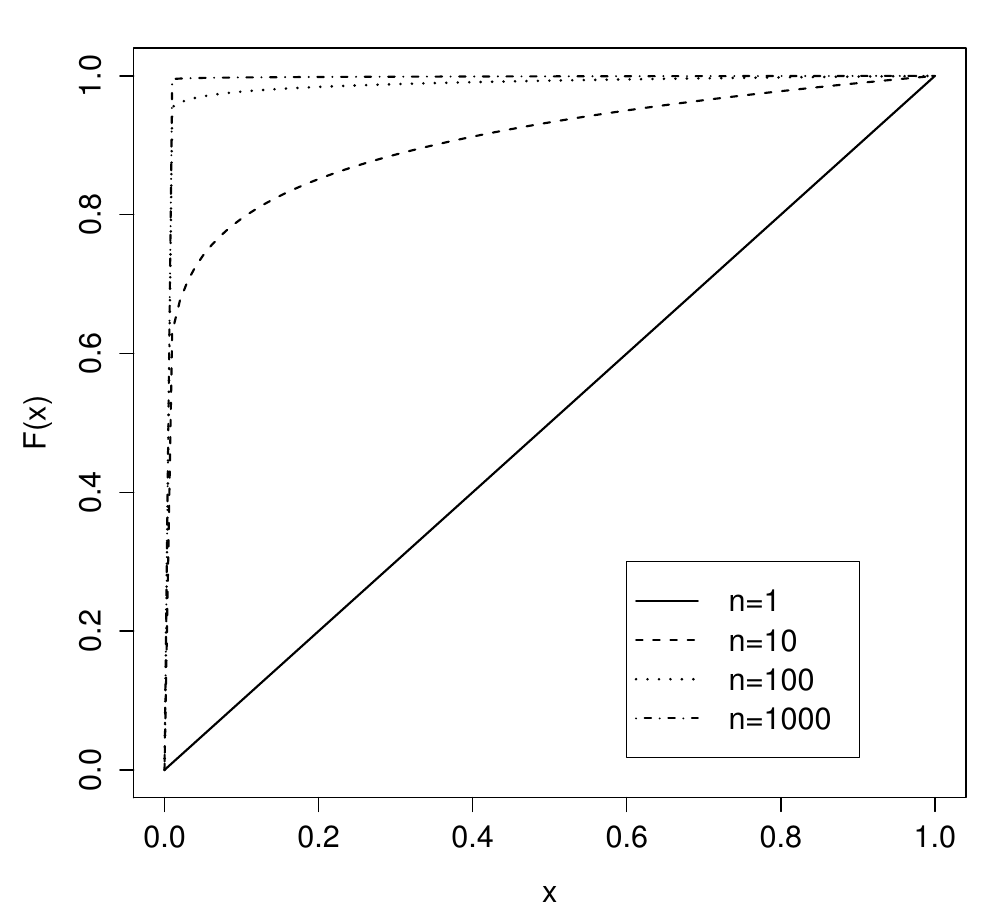}\\
  \caption{The ROC Curves Generated by $F(x)=x^{1/n}$.}
\end{figure}

We now repeat the above exercise by using the beta(2, 6) distribution as the ROC curve. The dinegentropy turns out to be 4.12548, which is greater than the dinegentropy generated by the ROC curve corresponding to the beta (1, 3) distribution. Thus we have here two distinct crossing ROC curves with a common AUC but different dinegentropies. Which of the two ROC curves corresponds to the better diagnostic test or a diagnostic test instrument?

To answer the above question, we will consider an ROC curve which, as a function of n, approaches the ideal ROC curve, and explore how its dinegentropy changes with n. To that effect, we consider the distribution function $F(x)=x^{\frac{1}{n}}$, $0\leq x\leq 1$. The probability density generated by $F(x)$ is $f(x)=\frac{1}{n}x^{-\frac{n-1}{n}}$. The behavior of $F(x)$ as a function of $n$ is shown in Figure 4.2. Note that as $n\rightarrow\infty$, $F(x)$ will approach the ideal ROC curve.

We now obtain the dinegentropy generated by $F(x)=x^{1/n}$, as a function of n. It turns out that for n=1, 10, 100, and 1000, the dinegentropy is 0, 11.685, 141.398, and 1439.814, respectively. Thus the dinegentropy for a random test is 0, and it increases with n, suggesting that diagnostic tests with larger dinegentropies are superior to those with smaller dinegentropies. In our example, the ROC curve based on the beta(2,6) distribution is therefore, superior to that based on the beta(1,3) distribution.

\subsubsection{Interpreting Dinegentropy}

What happens in the unlikely event that two separate diagnostic tests yield identical values for both the AUC and the dinegentropy? An obvious answer is that both the tests have the same efficacy. However, to investigate if this answer is the best we can hope for, we explore the geometrical structure of Theorem 4.1. We start by noting that the dinegentropy
\begin{eqnarray*}
\int_0^{\infty}[f_0(t)-f_1(t)]\text{log}_2 \left( \frac{f_0(t)}{f_1(t)} \right)dt.
\end{eqnarray*}\\[-20pt]
is a weighted average of the distance between $f_0(t)$ and $f_1(t)$ at $t$, weighted by the logarithm of the likelihood ratio at $t$. For $f_0(t)>f_1(t)$, the weight is positive, whereas for $f_0(t)<f_1(t)$, the weight is negative. This makes the dinegentropy a non-negative quantity. Furthermore, the larger the distance between $f_0(t)$ and $f_1(t)$, the larger is its assigned weight. Because of the concavity of the log-likelihood function, negative values of $[f_0(t)-f_1(t)]$ get accentuated more than the corresponding positive values, with the weights assigned to the latter almost attaining a plateau. Large negative values of $[f_0(t)-f_1(t)]$ are likely to occur when the spread of $f_1(t)$ is larger than the spread of $f_0(t)$; vice-versa for large positive values of $[f_0(t)-f_1(t)]$. This means that the dinegentropy as a measure of disparity tends to focus on the spreads of $f_0(t)$ and $f_1(t)$, with an emphasis on the spread of the latter. Since the assigned weights when $f_0(t)>f_1(t)$ are not symmetric with the assigned weights when $f_0(t)>f_1(t)$, and since $F_0$ and $F_1$ are assumed to be absolutely continuous, the only circumstance under which the dinegentropy of the two tests is equal is when $f_0(t)$ and $f_1(t)$ are the same for both tests.

It is of interest to compare the geometric structure of dinegentropy with that of the K-L distance when $F_1F_0^{-1}$ is the reference distribution, namely
\begin{eqnarray*}
\int_0^{\infty}f_0(t)\text{log}_2 \left( \frac{f_0(t)}{f_1(t)} \right)dt.
\end{eqnarray*}\\[-20pt]
Here we have a weighted average of $f_0(t)$ with the log-likelihood ratio as the weight. The weight assigned to $f_0(t)$ is positive when $f_0(t)>f_1(t)$; otherwise it is negative. The K-L construction therefore can lead to negative values of the distance metric; it can also lead to identical values of the metric for two diagnostic tests. The difference between the dinegentropy and the Kullback-Leibler distance metrics is that the former is a log-likelihood ratio weighted average of the differences between $f_0(t)$ and $f_1(t)$, whereas the latter is a weighted average of $f_0(t)$ alone. Given the elaborate construction underlying dinegentropy, it is literally impossible for two tests to have the same dinegentropy, other than under the case of identical tests.

\newpage


%
%

\end{document}